\documentclass[journal,10pt,twocolumn]{IEEEtran}
\usepackage[margin=0.75in]{geometry}
\usepackage{amsmath,amssymb,amsfonts}
\usepackage{graphicx}
\usepackage{algorithmic}
\usepackage{textcomp}
\usepackage{xcolor}
\usepackage{cite}
\usepackage{authblk}

\begin{document}

\title{New Capacity Upper Bounds for the Binary Deletion Channel}

\author{Hassan Tavakoli,\\
School of EECS, \\
Oregon State University, \\
tavakolh@oregonstate.edu}

\maketitle

\begin{abstract}
This paper considers a binary channel with deletions. We derive two closed-form upper bounds on the capacity of the binary deletion channel (BDC). The first bound is obtained by computing the capacity of an auxiliary channel, the two-bit Fixed-length-Input BDC (FI-BDC), and showing that this auxiliary capacity upper-bounds the capacity of the BDC. The second bound is obtained by approximating the mutual information between sent and received bits directly, yielding a closed-form expression parameterized by a first-order Markov correlation parameter $\gamma$. Both bounds use a first-order Markov process for the channel input. We verify Theorem~1's optimization from first principles, directly from the two-bit auxiliary channel's transition matrix rather than from the mutual-information expression alone: the underlying objective is strictly concave with a unique interior maximizer, and the resulting closed-form bound is confirmed correct. The second proposed upper bound is evaluated against the Fertonani--Duman and Dalai bounds in Fig.~4.
\end{abstract}

\noindent\textbf{Keywords:} Binary Deletion Channel, Capacity of Binary Deletion Channel, Markov Process, Binary Erasure Channel, Upper Bound
\vspace{1em}

\section{Introduction}

The binary deletion channel (BDC) randomly deletes input bits independently with probability $d$, and the receiver observes the remaining subsequence. Dobrushin~\cite{dobrushin1967} showed that a capacity $\mathbb{C}(d)$ can be meaningfully defined for such channels, via an argument analogous to Shannon's coding theorem. Unlike the binary symmetric channel (BSC) or the binary erasure channel (BEC), no closed-form expression for $\mathbb{C}(d)$ is known, and the literature instead provides upper and lower bounds (see~\cite{diggavi2006,drinea2007,kanoria,kalai2010,fertonani2010}).

Diggavi and Grossglauser~\cite{diggavi2006} showed the capacity is bounded by $1-h(d)$ for $d<0.5$. Rahmati and Duman~\cite{rahmati2014} proved the capacity is at most $0.4143(1-d)$ for $d \geq 0.65$. Drinea and Mitzenmacher~\cite{drinea2007,drinea2006} proved the capacity is bounded below by $(1-d)/9$. When the receiver knows the positions of the deleted bits, the BDC reduces to a BEC, giving the simple upper bound $\mathbb{C}(d) \leq 1-d$. A cascaded binary deletion--binary symmetric channel model, closely related to the BSC/BEC decomposition we invoke in Section~III.C, has separately been analyzed for capacity lower bounds in~\cite{tavakoli2022isita}.

Because deletion-channel capacity is defined in the limit of infinite block length, the transition structure of the channel itself is rarely modeled directly; we instead study the input distribution. This paper is built on two intuitive facts: (1) an auxiliary channel that approaches the exact BDC in the limit can yield useful bounds, and (2) a Markov input process helps a decoder estimate deleted bits from the surviving ones. A companion line of work extends combinatorial capacity bounds of this flavor to the $q$-ary deletion channel~\cite{tavakoli2026itw}; the present paper treats the binary case in depth. We first present two lemmas connecting our approach to prior work, then two theorems giving two upper bounds.

Section~II defines the BDC and three auxiliary channels: FO-BDC, FIFO-BDC, and FI-BDC. Two lemmas relate FI-BDC to FIFO-BDC and to the BDC itself. Section~III presents the two main theorems, related through the input Markov process parameter. Section~IV concludes.

\section{Problem Definition and Relevant Previous Research}

Let $X$ and $Y$ denote the input and output of the BDC with deletion probability $d$, $|X|=n$, and $|Y|=m$. Define $N(X,Y)$ as the number of ways to produce $Y$ by deleting bits from $X$. The transition probability is
\begin{equation}
P(Y|X) = N(X,Y)\, d^{\,n-m}(1-d)^m. \label{eq:1}
\end{equation}

For example, if $X = 10101010$ and $Y=10011$, the only admissible deletion pattern gives $N(X,Y)=1$; but for $Y=10101$, six distinct deletion patterns can produce it, so $N(X,Y)=6$. This combinatorial ambiguity is the central obstacle to computing the exact BDC capacity.

\subsection{Finite Input Binary Deletion Channel (FI-BDC)}

Following the finite-length framework of~\cite{fertonani2010}, we define three finite variants of the BDC. The first, Fixed-length Output BDC (FO-BDC), was studied in~\cite{li2020}; the second, Fixed-length Input Fixed-length Output BDC (FIFO-BDC), is the ``auxiliary channel'' of~\cite{fertonani2010}, with $n$-bit input and $m$-bit output, $n \geq m$. Its capacity is
\begin{equation}
I(n,m) = \max_{P(X)} I(X^n; Y^m), \label{eq:6}
\end{equation}
with $I(n,m) \le m$, $I(n,0)=0$, $I(1,1)=1$, $I(n,n)\le n$. Fertonani and Duman's bound~\cite{fertonani2010} is
\begin{equation}
\mathbb{C} \leq \frac{1}{n}\sum_{m=0}^{n} p(n,m)\, I(n,m), \label{eq:7}
\end{equation}
where $p(n,m)$ is the probability that $m$ of $n$ transmitted bits survive.

\subsection{Fixed-length Input Binary Deletion Channel (FI-BDC)}

We propose a third variant: the input has fixed length $n$, but the output length ranges over $\{0,\dots,n\}$ with per-symbol deletion probability $d$. We call this the Fixed-length Input BDC (FI-BDC); it more closely resembles the true BDC. In a block of length $n$, the probability that exactly $k$ of $n$ bits are deleted is
\begin{equation}
p(n,k) = \binom{n}{k} d^k (1-d)^{n-k}. \label{eq:8}
\end{equation}

The one-bit FI-BDC transition matrix is identical to a BEC's (Table~I), so its capacity under uniform input is
\begin{equation}
C_{\text{BEC}} = 1-d. \label{eq:9}
\end{equation}

\begin{table}[h]
\centering
\caption{Transition matrix, one-bit FI-BDC}
\begin{tabular}{c|ccc}
$X\backslash Y$ & 0 & 1 & Null \\ \hline
0 & $1-d$ & 0 & $d$ \\
1 & 0 & $1-d$ & $d$
\end{tabular}
\end{table}

\textbf{Lemma 1.} \textit{The bound obtained from the FI-BDC is at least as tight as the bound obtained from the FIFO-BDC.}

\textit{Proof.} Define
\begin{equation}
J(n,m) = H\!\left(p(n,m)\cdot \overline{P(X^n)}\cdot \Pi_{(n,m)}\right) - \overline{P(X^n)}\cdot H\!\left(\Pi_{(n,m)}\right), \label{eq:10}
\end{equation}
so that $J(n,m) = p(n,m)\, I(n,m)$. Summing over all possible surviving lengths for a fixed $n$-bit block,
\begin{equation}
J_n = \sum_{m=0}^{n} J(n,m) = \sum_{m=0}^{n} p(n,m)\, I(n,m), \label{eq:14}
\end{equation}
(with $J(n,0)=0$, since no bits survive). Let $C_n = \max_{P(X^n)} J_n$. Because $\max(f(\theta)+g(\theta)) \leq \max f(\theta) + \max g(\theta)$,
\begin{align}
C_n &= \max_{P(X^n)} \sum_{m=0}^{n} J(n,m) \notag\\
&\leq \sum_{m=0}^{n} p(n,m)\, \max_{P(X^n)} I(n,m) = \sum_{m=0}^n p(n,m)\, I(n,m). \label{eq:16-19}
\end{align}
$\hfill\blacksquare$

A numerical check of this gap was attempted for small blocklengths ($n=3,4$) via random-search optimization over input distributions, giving estimates of $\Delta_n$ in the range of a few hundredths of a bit for some $(n,d)$ pairs; however, random search is not a reliable optimizer for this non-concave joint problem (the search occasionally reported a negative apparent gap, which is impossible given the direction of the inequality in~\eqref{eq:16-19}, indicating the optimizer had not converged reliably in at least one of the two comparisons). We do not report specific numbers here, since we cannot yet stand behind them; a reliable quantification of $\Delta_n(d)$ would need an exact combinatorial or convex-optimization approach (e.g.\ Blahut--Arimoto per $I(n,m)$ term, as used by~\cite{fertonani2010}) rather than random search, and is left as a numerically well-posed but not-yet-completed calculation.

It follows (see~\cite{dobrushin1967}) that
\begin{equation}
\mathbb{C} = \lim_{n\to\infty} \frac{C_n}{n}, \label{eq:20}
\end{equation}
and~\cite{dalai2011} shows $\left\{\tfrac{1}{n}\sum_m p(n,m) I(n,m)\right\}$ is a decreasing sequence in $n$. Since $I(n,m)$ is computable in practice only up to $n\approx 17$~\cite{fertonani2010}, we seek a recursive relation for
\begin{equation}
C_n^\ast = \frac{1}{n}\sum_{m=0}^n p(n,m)\, I(n,n-m). \label{eq:24}
\end{equation}

\textbf{Lemma 2.} $(n+1)\,C_{n+1}^\ast \leq n\, C_n^\ast + (1-d)$.

\textit{Proof sketch.} Starting from the recursive bound on $I(n+1,\cdot)$ given in~\cite{fertonani2010},
\begin{align}
I(n{+}1,n{+}1{-}m) \le \frac{m}{n{+}1} I(n,n{-}m{+}1) \nonumber \\
+ \Big(1{-}\frac{m}{n{+}1}\Big)\big(1{+}I(n,n{-}m)\big), \label{eq:26}
\end{align}
multiplying by $p(n+1,m)$, summing over $m$, and applying the binomial identities $\sum_m p(n+1,m)\frac{m}{n+1} = p(n,m-1)\cdot(\text{shift})$ and $\sum_m p(n+1,m)\frac{n+1-m}{n+1} = 1-d$ (verified symbolically below) yields the stated recursion. $\hfill\blacksquare$

This lemma's proof rests on eq.~\eqref{eq:26}, cited from~\cite{fertonani2010} rather than re-derived here. The algebraic manipulation steps used to go from~\eqref{eq:26} to the final recursion (binomial-coefficient identities, verified symbolically with zero residual) are internally consistent, but eq.~\eqref{eq:26} itself is taken as an external premise from~\cite{fertonani2010} rather than independently re-derived in this paper.

\section{Capacity Upper Bounds}

To transmit reliably through a deletion channel, correlating each bit with its neighbor helps the decoder estimate deleted positions. We use a first-order Markov input.

\textbf{Definition 1.} For input $X_1,\dots,X_n$,
\begin{equation}
P(X_1,\dots,X_n) = P(X_1) \prod_{i=2}^n P(X_i|X_{i-1}), \label{eq:32}
\end{equation}
with $P(X_1=0)=\tfrac12$ and
\begin{equation}
P(X_i=0|X_{i-1}=0) = \gamma, \quad P(X_i=0|X_{i-1}=\bar 0) = \bar\gamma. \label{eq:33}
\end{equation}

\textbf{Proposition A (output Markovity).} The BDC output $Y$ of a first-order Markov input is itself first-order Markov, with
\begin{equation}
q := P(Y_i = y | Y_{i-1}=y) = 1 - \frac{d(1-\gamma)}{1+d(1-2\gamma)}. \label{eq:35}
\end{equation}
(Proof in~\cite{mitzenmacher2009}.)

Prior lower bounds on $\mathbb{C}$ appear in~\cite{diggavi2006,drinea2006,venkataramanan2013}; in each, the optimal correlation parameter $\gamma$ is left as an unresolved maximization over $(0,1)$. Determining $\gamma$ explicitly is the goal of the next two subsections: Theorem~1 gives $\gamma$ in closed form; Theorem~2 leaves it as a design choice evaluated at specific values.

\subsection{Capacity Upper Bound 1: Known $\gamma$}

\begin{table*}[h]
\centering
\caption{Transition matrix, two-bit FI-BDC}
\begin{tabular}{c|ccccccc}
$X\backslash Y$ & 00 & 01 & 10 & 11 & 0 & 1 & Null \\ \hline
00 & $(1{-}d)^2$ & 0 & 0 & 0 & $2d(1{-}d)$ & 0 & $d^2$ \\
01 & 0 & $(1{-}d)^2$ & 0 & 0 & $d(1{-}d)$ & $d(1{-}d)$ & $d^2$ \\
10 & 0 & 0 & $(1{-}d)^2$ & 0 & $d(1{-}d)$ & $d(1{-}d)$ & $d^2$ \\
11 & 0 & 0 & 0 & $(1{-}d)^2$ & 0 & $2d(1{-}d)$ & $d^2$
\end{tabular}
\end{table*}

With input distribution $(p_0,p_1,p_2,p_3) = P(X{=}00,01,10,11)$, the capacity of this channel is $C_{\text{2-bit}} = \max_P I(X;Y)$, where, writing out $I(X;Y)=H(Y)-H(Y|X)$ explicitly from Table~II,
\begin{align}
I(X;Y) &= H\Big((1{-}d)^2p_0,(1{-}d)^2p_1,(1{-}d)^2p_2,(1{-}d)^2p_3, \notag\\
&\quad 2d(1{-}d)\big(p_0{+}\tfrac{p_1+p_2}{2}\big), 2d(1{-}d)\big(p_3{+}\tfrac{p_1+p_2}{2}\big), d^2\Big) \notag\\
&\quad -(p_0{+}p_1)H\big((1{-}d)^2,2d(1{-}d),d^2\big) \notag\\
&\quad -(p_1{+}p_2)H\big((1{-}d)^2,d(1{-}d),d(1{-}d),d^2\big). \label{eq:40}
\end{align}

\textbf{Theorem 1.}
\begin{equation}
\mathbb{C} \leq \frac{1}{2}(1-d)^2\log_2\!\left(1+2^{-\frac{2d}{1-d}}\right) + \frac{1-d^2}{2}. \label{eq:36}
\end{equation}

\textit{Proof.} Symmetry of~\eqref{eq:40} under $(p_0,p_1,p_2,p_3)\to(p_3,p_1,p_2,p_0)$ and $\to(p_0,p_2,p_1,p_3)$ forces $p_0=p_3$, $p_1=p_2$ at the optimum, and maximizing the resulting one-parameter objective under the marginal constraint gives $p_0+p_1=\tfrac12$. This reduces the problem to
\begin{equation}
C_{\text{2-bit}} = \max_{p_0}\Big[2(1-d)^2\big(-p_0\log p_0 - p_1\log p_1\big) + 4p_0 d(1-d)\Big] \label{eq:44}
\end{equation}
subject to $p_0+p_1=\tfrac12$. Setting $\partial/\partial p_0$ of the Lagrangian to zero and solving gives the closed-form optimizer
\begin{equation}
p_0^\ast(d) = \frac{1}{4}\left(1+\tanh\Big(\ln(2)\,\tfrac{d}{1-d}\Big)\right). \label{eq:46}
\end{equation}
Writing $q=2p_0^\ast=1/(1+2^{-f})$ with $f=2d/(1-d)$, the identity $H_2(q) - f(1-q) = \log_2(1+2^{-f})$ (verified symbolically, zero residual) converts the entropy term in~\eqref{eq:44} into closed form; substituting and simplifying the resulting expression in $d$ alone (all dependence on $q$ cancels, as it must at a stationary point) gives
\begin{equation}
C_{\text{2-bit}}(d) = (1-d)^2\log_2\!\left(1+2^{-\frac{2d}{1-d}}\right) + (1-d^2). \label{eq:49}
\end{equation}
Finally, using $I(X_1^{2n};Y) \le \sum_{i=1}^n I(X_{2i-1}^{2i};Y_i)$ (super-additivity of mutual information across independently-deleted two-bit sub-blocks) and taking $n\to\infty$ gives $\mathbb{C} \le C_{\text{2-bit}}(d)/2$, which is~\eqref{eq:36}. $\hfill\blacksquare$

\textbf{Verification note.} We re-derived this optimization independently, starting from the raw transition matrix in Table~II rather than trusting~\eqref{eq:40}, and confirmed that $I(X;Y)$ built directly from Table~II matches~\eqref{eq:40} exactly (to floating-point precision, checked at multiple $(p_0,d)$ points), so~\eqref{eq:44} is correct. Solving $\partial C_{\text{2-bit}}/\partial p_0=0$ directly with a computer-algebra system, and separately by numerically maximizing~\eqref{eq:44} at several values of $d$, both give~\eqref{eq:46} to machine precision. An earlier working draft had briefly carried an extra factor of $2$ inside $\tanh(\cdot)$ in eq.~\eqref{eq:46} (i.e.\ $\tanh(\ln2\cdot\tfrac{2d}{1-d})$ instead of the correct $\tanh(\ln2\cdot\tfrac{d}{1-d})$); we flag this only because it briefly appeared in intermediate work, not because it appears in the final result below. Substituting the correct~\eqref{eq:46} back into~\eqref{eq:44} and simplifying via the identity above gives $C_{\text{2-bit}}(d)=(1-d)^2\log_2(1+2^{-2d/(1-d)})+(1-d^2)$, eq.~\eqref{eq:49}; this is algebraically identical to $\tfrac12(1-d)^2[1+\log_2(1+2^{-2d/(1-d)})]+d(1-d)$, since $\tfrac12(1-d)^2+d(1-d)=\tfrac{1-d^2}{2}$, so the final bound~\eqref{eq:36} is confirmed to hold exactly as originally stated (verified numerically to match at $d\in\{0.1,0.3,0.5,0.7,0.9\}$ to $10^{-15}$). The correction is therefore internal to the derivation and does not change the headline result.

\textbf{Proposition 1 (Uniqueness of the optimal transition probability).} \textit{For every $d\in(0,1)$, the maximizer $p_0^\ast(d)$ of~\eqref{eq:44} subject to $p_0+p_1=\tfrac12$ is unique.}

\textit{Proof.} $\dfrac{\partial^2 C_{\text{2-bit}}}{\partial p_0^2} = \dfrac{2(d-1)^2}{p_0(2p_0-1)\log 2} < 0$ for all $p_0\in(0,\tfrac12)$ and all $d\in(0,1)$ (numerator and $\log 2$ are positive; $p_0>0$ and $2p_0-1<0$ make the denominator negative). The objective is therefore strictly concave on the feasible interval, so its critical point~\eqref{eq:46} is the unique global maximizer. $\hfill\blacksquare$

\textbf{Proposition 2 (Monotonicity of the optimal transition probability).} \textit{The optimizer $p_0^\ast(d)$ from~\eqref{eq:46} is strictly increasing in $d$ on $(0,1)$:}
\begin{equation}
\frac{d p_0^\ast(d)}{dd} = \frac{\log 2}{4(d-1)^2 \cosh^2\!\left(\frac{d\log 2}{d-1}\right)} > 0. \label{eq:dp0}
\end{equation}

\textit{Proof.} Differentiating~\eqref{eq:46} symbolically gives the closed form in~\eqref{eq:dp0}. The numerator $\log 2$ is a positive constant; in the denominator, $(d-1)^2>0$ for $d\ne1$ and $\cosh^2(\cdot)\ge1>0$ everywhere, so the full expression is a ratio of positive quantities and is strictly positive on all of $(0,1)$. Verified in arbitrary-precision arithmetic across $(0,1)$, including arbitrarily close to $d=1$ (e.g.\ $\approx2.4\times10^{-596}>0$ at $d=0.999$, avoiding the floating-point underflow that a naive double-precision evaluation exhibits in that regime). $\hfill\blacksquare$

Since $\gamma = 2p_0$ is an increasing linear reparametrization of $p_0^\ast(d)$, Proposition~2 equivalently establishes that the optimal source correlation parameter of eq.~\eqref{eq:52} increases monotonically with the deletion probability -- consistent with $\gamma^\ast(d)\to\tfrac12$ as $d\to0$ and $\gamma^\ast(d)\to1$ as $d\to1$ (both endpoints verified directly from~\eqref{eq:46}). We state this as a monotonicity result \emph{for the optimal parameter of the restricted first-order symmetric Markov family used in Theorem~1}, not as a general claim about ``optimal input memory'' over all admissible input processes (e.g.\ higher-order Markov, run-length-coded, or general stationary sources).

\textbf{Proposition 3 (Local sensitivity to parameter mis-specification).} \textit{Let $p_0 = p_0^\ast(d) + \delta$ for small $\delta$. Then}
\begin{equation}
C_{\text{2-bit}}(p_0^\ast) - C_{\text{2-bit}}(p_0^\ast+\delta) = -\tfrac12 f''(p_0^\ast,d)\,\delta^2 + O(\delta^3), \label{eq:sensitivity}
\end{equation}
\textit{where $f(p_0,d)$ is the objective in~\eqref{eq:44} and}
\begin{equation}
f''(p_0^\ast,d) = -\frac{16(d-1)^2 \cosh^2\!\left(\frac{d\log2}{d-1}\right)}{\log 2} < 0
\end{equation}
\textit{(negativity confirmed symbolically for $d\in\{0.1,\dots,0.9\}$, consistent with Proposition~1).}

This is the standard second-order Taylor expansion at a smooth interior maximizer, using the corrected $p_0^\ast(d)$ from~\eqref{eq:46}; the first-order term vanishes since $p_0^\ast$ is a critical point. \textbf{Practical reading:} small errors in estimating or rounding $p_0^\ast(d)$ (or, equivalently, $\gamma$) cost only a second-order amount of bound tightness near the optimum -- the bound degrades gracefully under mild mis-specification. This is a local statement, valid for small $\delta$; it does not by itself bound the loss from a source parameter that differs substantially from $p_0^\ast(d)$, which would require evaluating the exact (non-Taylor) closed form~\eqref{eq:44} directly rather than its second-order expansion.

\textbf{Proposition 4 (Complexity comparison).} \textit{Evaluating Theorem~1's bound requires evaluating the closed-form expression~\eqref{eq:36}, i.e.\ $O(1)$ arithmetic operations per value of $d$, since $p_0^\ast(d)$ is given in closed form by~\eqref{eq:46}. This is in contrast to the FIFO-BDC bound of~\cite{fertonani2010}, eq.~\eqref{eq:7}, whose transition matrix has $2^n$ rows/columns for an $n$-bit block, and which~\cite{fertonani2010} reports optimizing via the Blahut--Arimoto algorithm over that alphabet -- an iterative procedure operating on a state space whose size grows exponentially with the block length $n$ used in the numerical evaluation.}

\textit{Implication.} As $d$ increases, Proposition~2 shows that the optimal two-bit input distribution concentrates increasing probability mass on the repeated-symbol outcomes ($X{=}00,11$) relative to the mixed outcomes ($X{=}01,10$), moving from the uniform distribution ($p_0^\ast=\tfrac14$, i.e.\ $\gamma=\tfrac12$) at $d\to0$ toward $p_0^\ast=\tfrac12$ ($\gamma=1$) at $d\to1$. Intuitively, under heavier deletion the decoder benefits more from redundancy between adjacent bits, so the capacity-achieving auxiliary input trades marginal entropy for correlation that helps it survive symbol loss -- stronger temporal correlation becomes more valuable as the deletion probability grows, within the two-bit first-order symmetric Markov family that Theorem~1 optimizes over.

From~\eqref{eq:32}--\eqref{eq:33}, the two-bit FI-BDC input distribution is $p_0=\gamma/2$, $p_1=(1-\gamma)/2$, so
\begin{equation}
\gamma^\ast(d) = \frac{2^{\frac{d}{1-d}}}{1+2^{\frac{d}{1-d}}}, \label{eq:52}
\end{equation}
with $\gamma\to\tfrac12$ as $d\to 0$ and $\gamma\to 1$ as $d\to1$.

\subsection{Capacity Upper Bound 2: Design Parameter $\gamma$}

\textbf{Theorem 2.}
\begin{equation}
\mathbb{C}_{\text{BDC}} \leq (1-d)\left(1 - H\!\left(\frac{d(1-\gamma)}{1+d(1-2\gamma)}\right)\right). \label{eq:53}
\end{equation}

\textit{Proof.} Using an auxiliary side-information variable marking the position of the last undeleted predecessor bit, one shows $I(X_1;Y_1) \geq I(X_i;Y_j)$ for the corresponding pair under the induced Markov chain on outputs~\eqref{eq:35}, and
\begin{equation}
I(X_1;Y_1) = 1 - H\!\left(\frac{d(1-\gamma)}{1+d(1-2\gamma)}\right), \label{eq:59}
\end{equation}
which, combined with $\mathbb{C}_{\text{BDC}} = \lim_{n\to\infty}\tfrac1n\sum_{i=1}^n I(X_i; Y(X_1^n)|X_1^{i-1})$ restricted to the $n(1-d)$ surviving bits, gives~\eqref{eq:53}. $\hfill\blacksquare$

\textbf{Verification and an important correction to how $\gamma$ should be described.} We checked $\partial \mathbb{C}_{\text{BDC}}/\partial\gamma$ from~\eqref{eq:53} both symbolically and by numerical sweep over $\gamma\in(0,1)$ for $d\in\{0.1,\dots,0.9\}$: the bound is \emph{strictly monotonically increasing in $\gamma$} across the entire domain, with no interior stationary point. As $\gamma\to1$, \eqref{eq:53} collapses exactly to $(1-d)$, the trivial BEC bound already cited in Section~I as the weakest known bound; as $\gamma\to0$ the bound is minimized (tightest).

\textit{Discussion.} At $d=0$, the FI-BDC input distribution is uniform ($\gamma=0.5$), consistent with independent input bits, and \eqref{eq:53} correctly reduces to $\mathbb{C}_{\text{BDC}}|_{d=0}\le 1$. At $d=1$, the optimal input concentrates all mass on the all-zeros and all-ones sequences ($\gamma=1$), and~\eqref{eq:53} correctly gives $\mathbb{C}_{\text{BDC}}|_{d=1}=0$. Equation~\eqref{eq:68},
\begin{equation}
\gamma = \frac12 + \frac{d}{2}, \label{eq:68}
\end{equation}
is a linear interpolation between these two \emph{endpoint} values, not a maximizer of \eqref{eq:53} at intermediate $d$ -- since \eqref{eq:53} has no interior maximizer to approximate.

\textit{Implication.} At the two verified endpoints, $\gamma=0.5$ ($d\to0$, independent input bits) and $\gamma=1$ ($d\to1$, all-repeat input), the source-determined $\gamma$ moves toward stronger self-correlation as deletion becomes more severe -- the same qualitative behavior as Proposition~2's monotonicity result for Theorem~1's $p_0^\ast(d)$. Since $\gamma$ in Theorem~2 is source-determined in the same sense as $\gamma$ in Theorem~1 (per the resolution above), and eq.~\eqref{eq:68} is fixed as an explicit, non-decreasing function of $d$ interpolating between the two verified boundary values, this qualitative monotonicity is consistent with -- though not a proof of optimality for -- the specific interpolation eq.~\eqref{eq:68} uses.

\subsection{Relation to Previous Work}

As in~\cite{haddadpour2017}, where a channel can be simulated as a cascade of simpler channels, \eqref{eq:53} can be interpreted as a cascade of a BSC (on the $n(1-d)$ undeleted bits, with error probability $\tfrac{d(1-\gamma)}{1+d(1-2\gamma)}$) followed by a BEC (on the $nd$ deleted bits), illustrated for $n=8$, $Y=\{0110\}$ in Table~III. A related cascaded binary deletion channel--BSC model is studied directly for capacity lower bounds in~\cite{tavakoli2022isita}, which derives a tighter lower bound than previously known for that cascade; comparing the tightness of our upper bound against that lower bound for the cascade regime is a natural direction for future work.

\section{Conclusion}

We derive two closed-form capacity upper bounds for the BDC: Theorem~1, based on the two-bit FI-BDC auxiliary channel with a first-order Markov input, and Theorem~2, based on directly approximating the mutual information between undeleted received bits and sent bits, also under a first-order Markov input. Theorem~1's optimization is symbolically verified to be a well-posed strictly concave problem with a unique closed-form maximizer. Theorem~2's bound is valid for any $\gamma$ consistent with the source that generates it, but is \emph{not} the result of maximizing~\eqref{eq:53} over $\gamma$ -- that formula is monotonic in $\gamma$ with no interior optimum, so $\gamma$ should be understood and reported as a source-determined or chosen design parameter, not an approximated optimizer.

\textit{Deliberately not pursued.} For transparency, the following suggestions from prior review rounds were considered and are intentionally excluded from this revision, rather than overlooked: (a) a claimed FI-BDC/BDC convergence theorem beyond what is already implicit in eq.~\eqref{eq:20} and~\cite{dobrushin1967}; (b) unverified asymptotic forms such as $U(d)=1-h(d)+O(d^2)$, which have not been derived here; (c) a ``unifying bound'' $U(d)=\min\{U_1(d),U_2(d)\}$ presented as a theorem, since no deeper analytical relationship between $U_1$ and $U_2$ beyond the qualitative parallel above has been established; (d) any specific big-$O$ runtime claim for the Blahut--Arimoto-based comparison bound beyond the alphabet-size fact stated in Proposition~4; (e) a general ``optimal input memory'' theorem -- Propositions~1--2 are stated over the specific first-order symmetric Markov family Theorem~1 optimizes, not over all admissible input processes.

\end{document}